\documentclass[prl,floatfix,twocolumn,showkeys]{revtex4-1}

\usepackage{color} 
\usepackage{graphicx} 
\usepackage{amsmath,amssymb} 

\newcommand{\be}{\begin{equation}}
\newcommand{\ee}{\end{equation}}

\begin{document}

\author{Stefano Mostarda}
\author{Federico Levi}
\author{Diego Prada-Gracia}
\author{Florian Mintert}
\email{florian.mintert@frias.uni-freiburg.de}
\author{Francesco Rao}
\email{francesco.rao@frias.uni-freiburg.de}

\affiliation{Freiburg Institute for Advanced Studies, School of Soft
Matter Research, Albert-Ludwigs Universitaet Freiburg, Albertstrasse 19, Freiburg im Breisgau, 79104, Germany.}

\title {Structure-dynamics relationship in coherent transport through disordered systems}

\begin{abstract} 

 Quantum transport is strongly influenced by interference with phase relations
 that depend sensitively on the scattering medium.
  Since even small changes in the geometry of the medium can turn constructive
  interference to destructive, a clear relation between structure and fast,
  efficient transport is difficult to identify. Here we present a complex
  network analysis of quantum transport through disordered systems to elucidate
  the relationship between transport efficiency and structural organization.  Evidence is
  provided for the emergence of structural classes with different geometries
  but similar high efficiency.  Specifically, a structural motif characterised by
  pair sites which are not actively participating to the dynamics 
  renders transport properties robust against perturbations. Our results pave the way for a systematic rationalization of the design
  principles behind highly efficient transport which is of paramount importance for technological applications as well as to address transport robustness in natural light harvesting complexes.  
\end{abstract}

\date{\today}

\maketitle

\section{Introduction}
Transport of charge or energy through disordered landscapes is one of
the most fundamental mechanisms underlying biological and technological
functionality \cite{Scholes2011, coropceanu2007charge, pivrikas2007review}.  If
the entities that are being transported behave wave-like, {\it i.e.} propagate
coherently, interference resulting from scattering off the disordered medium
can result in strong focusing behavior due to constructive interference, as
observed for example for an electron gas \cite{nature.410.183.2001}, the
coherent back-scattering of light from atomic clouds \cite{labeyrie99} and
predicted for Bose-Einstein condensates \cite{PhysRevLett.109.190601}.  When
this focus lies in the region to which an object should be transmitted,
coherent behavior results in enhanced transport as compared to incoherent (particle like)
processes \cite{leegwater1996coherent, chachisvilis1997excitons}.
Consequently, it would be highly desirable to exploit such enhanced transport
mechanisms.

Constructive interference, however, relies on well-defined phase relations that
need to be satisfied rather accurately, but get easily altered due to
small changes in the geometry of the scattering medium.  The onset of
destructive interference then reduces transport, or suppresses it completely
\cite{PhysRev.109.1492}.  This results in a highly complicated
structure-functionality relationship: two structures with hardly noticeable
geometric differences can lead to strongly different transport properties and
two structures with similar transport properties might not share any geometric
similarity.  It is thus rather hard to identify geometric features associated
with good transport, what would be absolutely necessary for the use
of constructive interference as design principle in technological applications.

The inherent complexity of the problem as well as the large number of degrees
of freedom involved calls for the application of advanced statistical tools.
Inspired by the substantial achievements of network science to elucidate
complex systems like for instance economic growth \cite{hidalgo2007product},
human diseases \cite{goh2007human} and organic chemistry
\cite{grzybowski2009wired}, our aim is to shed light on the elusive
relationship between the structure of disordered media and constructive
interference through the application of a network approach.

This allows to identify a clear structural motif formed by pair sites that are not actively involved in the dynamics, but provide both enhancement of transport and robustness against random displacement of the sites. Our results can be used as a starting point to address robustness of transport in natural systems like light harvesting complexes.

\section{Results}
\subsection{The quantum transport model}
We consider a discrete two-level $N$-body
system, whose interactions are described by a tight-binding Hamiltonian
\begin{equation} H=\sum_{i\neq j}^N J \frac{r^{3}_{0}}{|\vec {\bf r}_i-\vec
{\bf r}_j|^3}\sigma_i^{-}\sigma_j^{+}\ ,
\label{eq:ham}
 \end{equation} 
where $\sigma_i^{-/+}$ describe the annihilation/creation of an excitation at
site $i$, $J$ is the coupling constant and $r_0$ defines the natural length scale of the system.  The interaction rate decays
cubically with the distance between the sites in accordance with dipole-dipole
interaction.  Within this model a \emph{structure} is defined by the positions
of the $N$ sites.  The initially excited site (input) and the output site where the excitation has to be delivered are located at
diagonally opposite corners of a cube of side $r_0$, while the remaining $N-2$
sites are placed at random positions within this cube.  

While we looked at systems with $N$ ranging from 4 to 8, most of our attention focused
on $N=6$, being the smallest set for which non-trivial behavior emerged.  For this case,
a large sample of 100 millions random structures was generated. 
This sample covered the whole spectrum of transport efficiencies $\epsilon$, 
where $\epsilon$ is defined as the maximal probability
\begin{equation} 
\epsilon=\mathrm{max}_{t\in[0,\tau]}|\langle \text{in} |\mathrm{e}^{iHt}| \text{out} \rangle|^2
\label{eq:emaxtau}
\end{equation}
to find the excitation at the output site $| \text{out} \rangle$ within a short time interval after initialization.
In order to make sure that we exclusively target fast transport that necessarily needs to result from constructive interference, we chose 
$\tau= \frac{1}{10} \frac{2\pi \hbar}{J} \frac{r_{\text{in-out}}^3}{r_{0}^3}$, {\it i.e.} 
a time-scale that is ten times shorter than the scale associated with direct interaction between input and output sites \cite{scholak2011} (see Supplementary Note 1 for further details).

Being interested in the characterization of efficient transport, our analysis
focused on structures with $\epsilon>0.9$, a property that is satisfied by
only $14280$ configurations out of the generated $10^{8}$.  From a structural point
of view this reduced set is highly \emph{heterogeneous}, meaning that two
structures with similar efficiency do not necessarily share any evident common pattern
\cite{scholak2011}. Such structural variety hinders a straightforward
interpretation of the geometrical features compatible with efficient transport.

\begin{figure}[t]
  \includegraphics[width=90mm]{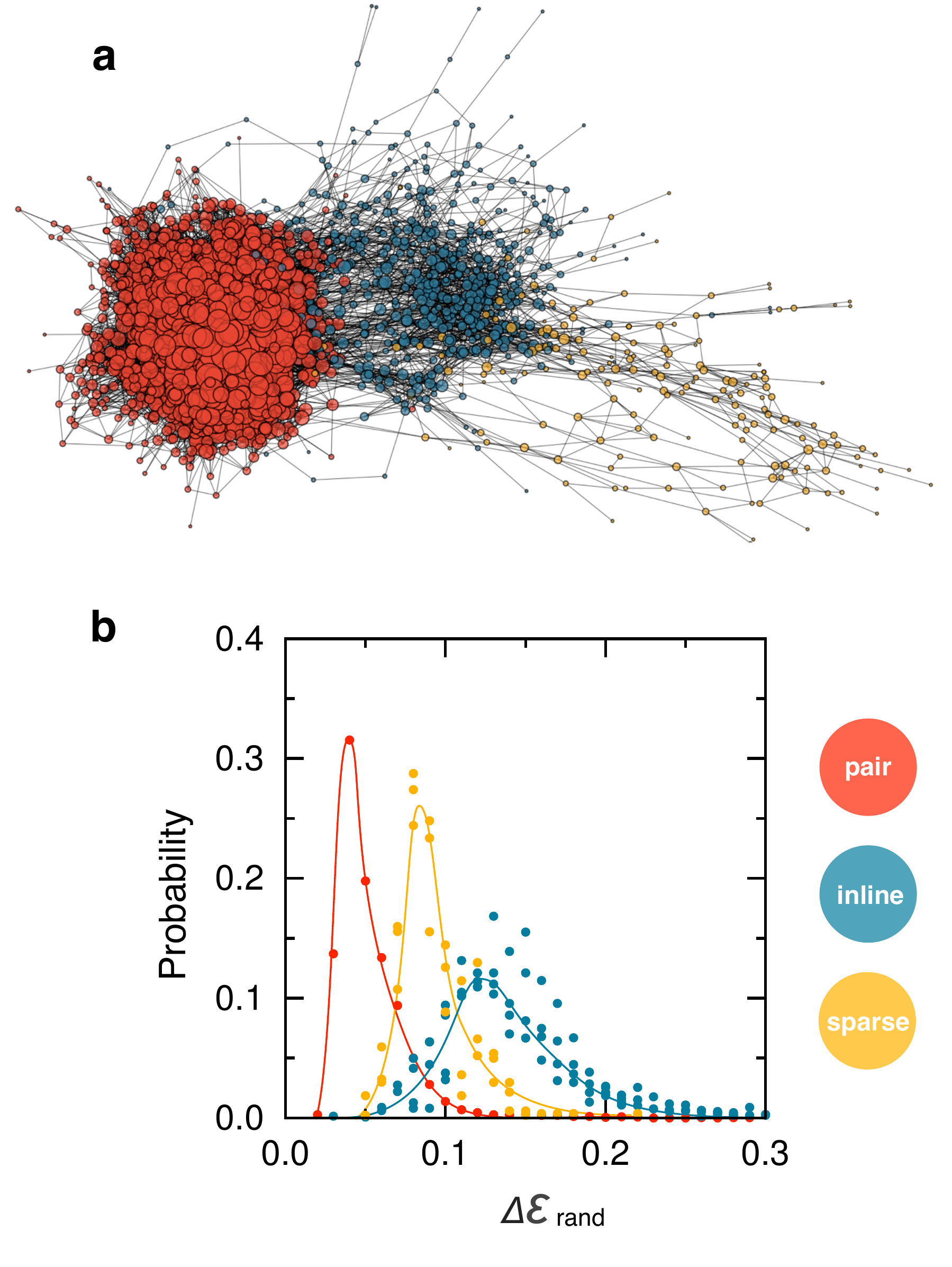}
  \caption{The quantum efficiency network. (a) Nodes in the network represent
    structures with six excitable sites ($N=6$) and $\epsilon>0.9$. Links are placed
    if two structures are geometrically similar. The
    network layout is obtained via the Fruchterman-Reingold algorithm
    \cite{fruchterman1991} which puts nodes with several neighbors in common
    close in space. To avoid overcrowding in the network layout, only 1/5
    of the nodes have been represented in the picture. Node size is proportional to the number
    of links. (b) The probability distribution of efficiency loss upon random displacements
    for the eight clusters (calculated on the whole sample, see main text for details). All eight distributions are
    shown where colors are chosen to highlight the presence of three classes (lines are
    a guide to the eye).  Network nodes color
    coding follows the class definitions: pair (red), inline (blue) and sparse
  (yellow).}
  \label{fig:network}
\end{figure} 

\subsection{The quantum efficiency network}
To unravel this structure-dynamics relationship, we applied a set of tools
based on complex networks originally developed for the characterization of
molecular systems \cite{rao2004,gfeller2007}.  Specifically, these methods are designed to
analyze large ensembles of configurations, potentially allowing in this case a systematic
classification of structures which lead to exceptional transport.
We generated a complex network where the 14280 structures with $\epsilon>0.9$
represent the nodes and a link is placed between them if two structures are
geometrically similar independently on the specific dynamics of the excitation.  The parameter used to estimate structural
similarity is $S^2 = \sum^N_{i=1}  d_i ^2/N$, where $d_i \, , \, i\in
\{1,...,6\}$ is the distance between corresponding sites in any two structures.
  Links were placed when $S^2<0.0125 \ r^2_0$ (see Methods for further details on the network creation protocol). The resulting
quantum efficiency network is shown in Fig.~\ref{fig:network}a.  In this
picture, nodes are proportionally close in space according to the amount of
common neighbors. The color coding adopted for this network will be discussed
in detail below, anticipating at this stage that the presence of densely
connected regions, i.e. clusters of nodes which are highly interconnected among
each other, indicates the presence of groups of structures with common geometrical motifs.

We identified these regions using a network clusterization algorithm based on a
self-consistency criterion in terms of network random walks \cite{gfeller2007, Van2008}
(see Methods) that split the network into eight
clusters comprised of structures with similar sites arrangements.
  The eight clusters have very different relative populations,
respectively of 73.3\%, 9.2\%, 4.7\%, 3.2\%, 2.6\%, 1.9\%, 1.2\%, 1.1\%.  As
such, there is a bias towards certain structure types with respect to others.
Cumulatively the eight clusters represent the 97\% of the whole sample with an
unclassified 3\% due to noise (see Supplementary Figure S1).

\begin{figure*}[t]
  \includegraphics[width=120mm]{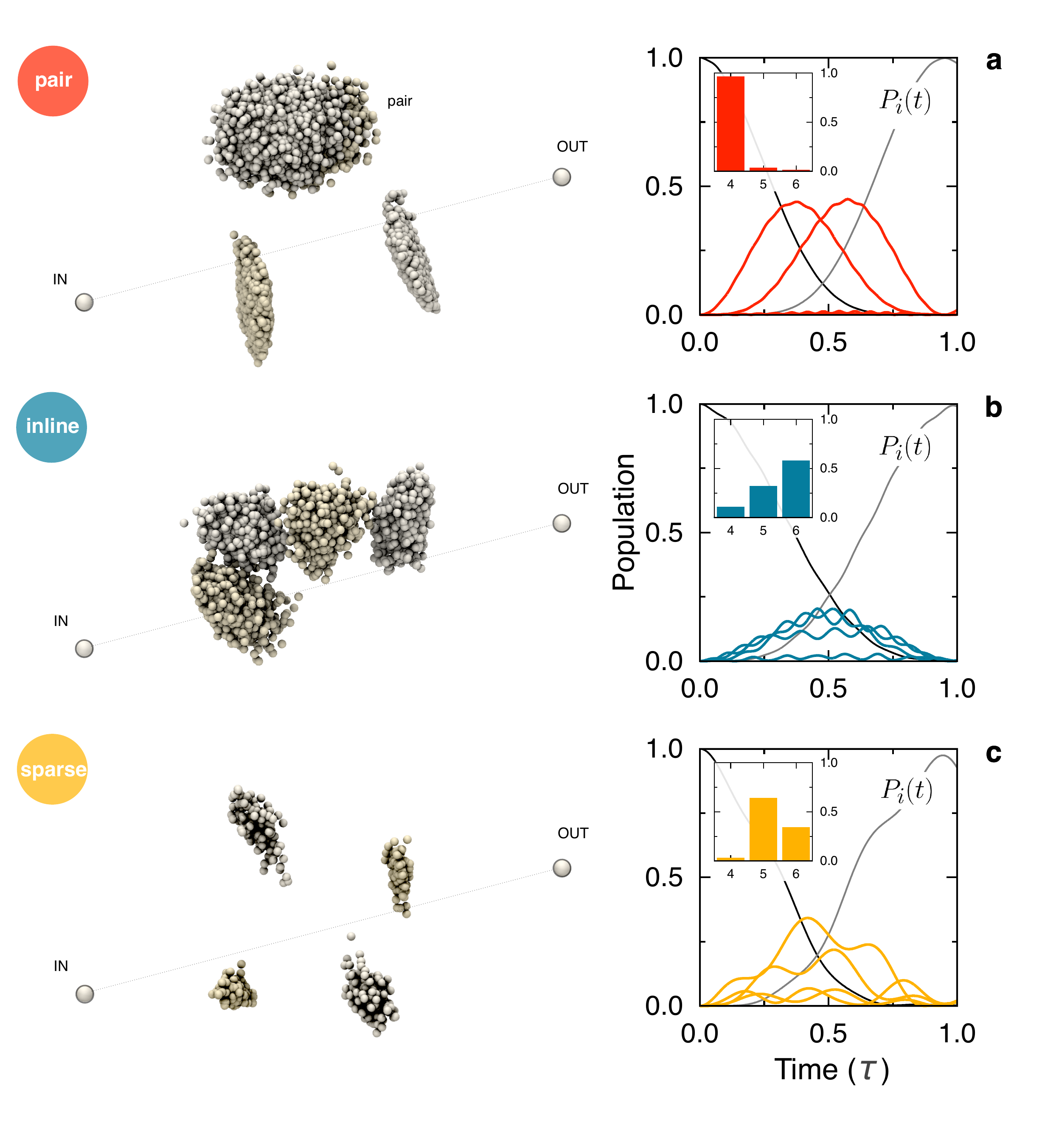}
  \caption{Structure-dynamics relationship for the (a) pair, (b) inline and (c) sparse 
    classes. On the left side of the figure structures belonging to the cluster
    containing the most efficient configuration of the class are overimposed (namely cluster
    number 1, 2 and 7 respectively, see Supplementary Figures S2 ans S3
    for the remaining ones). Characteristic structural motifs emerge from each of the
    different clusters.  The exciton dynamics of the most efficient
    structure of each class is shown in the right part of the figure where the
    x and y-axis represent the time and the excitation probability (population),
    respectively. Input and output sites are shown as black and dark grey lines. The relative frequencies to find structures with four, five or
    six active sites (including input and output sites) within each class are shown as insets (see main text for details). Sites
    are colored alternatively in light and dark grey in order to highlight their different position. Structural 
    rendering was done with VMD \cite{humphrey1996vmd}.} 
  \label{fig:dynamics}
\end{figure*}

\subsection{Efficiency loss upon random displacement}
The detection of eight clusters does not imply the presence of eight distinct
dynamical behaviors.  We therefore probed transport robustness against random
displacements of the individual sites of a structure, with displacements
restricted to a cube of side $0.05 \  r_0$ centered around the original position
of the site.  For each structure, $\mathit{\Delta} \epsilon_{\text{rand}}$ was calculated as the
original efficiency minus the average efficiency obtained from 1000
site-randomizations. In this scheme, structures  were kept rigid under the
assumption that the dynamics occurs on a much faster time scale than
low-frequency fluctuations of the entire system (e.g.  in the context of
biological systems this would be equivalent to large-scale protein breathing).
The distributions of $\mathit{\Delta} \epsilon_{\text{rand}}$ for the eight clusters are shown in
Fig.~\ref{fig:network}b. Strikingly, the data spontaneously grouped into three distributions 
that are colored in red, blue and yellow in the figure. 
This coding split the network in homogeneously colored parts, as shown in Fig.~\ref{fig:network}a. This was a-priori not obvious
as the network could have shown a certain degree of color mixing,
providing strong evidence that network topology and excitation dynamics are
correlated properties.  Summarizing, the random
displacements analysis provided evidence that (i) all structures within a
cluster show similar response upon perturbation and (ii) three well-defined
types of responses emerge from the eight clusters.

\subsection{Classes of quantum behavior}
These results suggested a characterization of the whole sample of efficient structures into three
classes of similar quantum behavior. The classes are named {\it pair},
{\it inline} and {\it sparse} because of their average geometrical properties.  In
fact, they respectively show couples of sites very close to each other, a
compact arrangement around the input/output axis and a more sparse geometry.
In terms of clusters, the pair class includes only one cluster (the most
populated one, 73.3\% of the whole sample) while the inline includes four
clusters (17.6\%) and the sparse only three (6.3\%).  

\begin{figure}[t]
  \includegraphics[width=85mm]{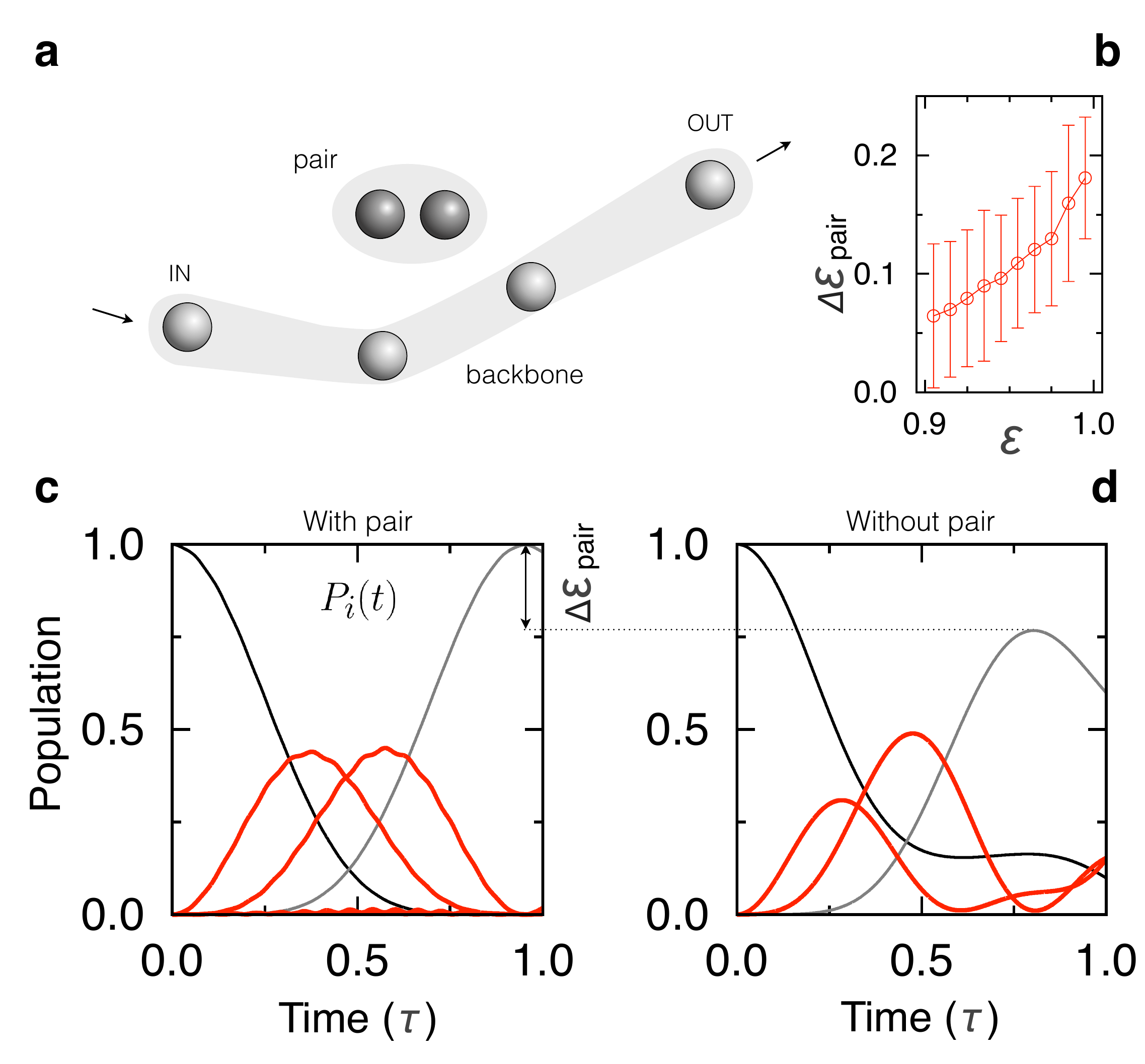} 
  \caption{Characterization of the pair class. (a) The most efficient structure: pair
    and backbone sites are shown in dark and light gray, respectively.  (b)
    Efficiency loss upon pair removal as a function of the original efficiency $\epsilon$. Error bars are calculated according to
    the standard deviation. (c-d) Comparison of the typical dynamics with and without
    the pairs. The coherent signal does not focus in the output site in the absence of the
    pair sites, resulting in an efficiency loss $\mathit{\Delta} \epsilon_{\text{pair}}$ of up to 0.32 
    (in the plot the dynamics of the most efficient structure is shown, where $\mathit{\Delta} \epsilon_{\text{pair}}=0.23$).}
  \label{fig:pairs}
\end{figure}

The pair class is very robust against random displacements, with an average loss of
efficiency of around 0.06 (red data in Fig.~\ref{fig:network}b). Interestingly, this
class performed much better than the inline and sparse classes which showed an
average loss of 0.14 and 0.10, respectively (with losses up to 0.3 for the
former). These results indicated that the geometry in these latter
classes is more correlated while in the pair class sites can be moved by
small displacements in an independent fashion.  However, robustness comes with a
price: the pair class is generally slower. For the fastest processes in each class, the
times at which the population in the output site is maximal are 
0.67, 0.30 and 0.54 $\tau$ for the pair, inline and sparse class,
respectively, while in average these values are of 0.92, 0.83 and 0.86 $\tau$ for the three classes, respectively.

In Fig.~\ref{fig:dynamics} several properties of the three classes are illustrated. On the left side of the picture
structures belonging to the cluster with highest efficiency within the class are overimposed on top of each other (see Supplementary Figures S2 and S3 
for the remaining ones). This representation allows a visual appreciation of the structural homogeneity within a cluster as well as
of the diversity among clusters. In the right part of the picture, the exciton
dynamics of the most efficient structure of the class is shown.  In all
three cases transport efficiency is larger than 0.97.  However, the three dynamics
differs substantially. In the pair class (right part of Fig.~\ref{fig:dynamics}a), two of
the intermediate sites are successively excited with no active role of the
remaining other sites (excluding input and output). Conversely, the other two classes show more complex
patterns of excitation. These results provide evidence for a strong
structure-dynamics relationship given that the final values of the efficiency
are very similar in all three cases.  

\subsection{Inactive pair sites enhance transport}
The pair class shows a prototypical modular structure.  The first
module is comprised of four sites including the input and output
approximately lined up along the input/output axis, defining a
\emph{backbone} for the entire structure (light gray spheres in Fig.~\ref{fig:pairs}a).  The second module is formed by the
remaining two sites, the \emph{pair} (dark gray spheres).  Backbone sites are
approximately equally spaced between input and output with typical
inter-sites distances of around $0.60-0.64 \ r_0$. 
Pair sites instead are always very close to each other with a
inter-site distance of around 
$0.25 \ r_0$  (see Supplementary Figure S4).

The position of the pair is more heterogeneous than the backbone sites, i.e.
their position in space changes between structures as indicated by the disperse
cloud of sites in the structural overlaps of Fig.~\ref{fig:dynamics}a. A first indicator
on the origin of the increased robustness is then given: pair sites can be moved within a larger volume
without dramatically affecting the transport efficiency (Supplementary Figure S4).

Backbone and pair sites show a completely
different dynamical behavior.  Systematic analysis of the distribution of the
maximum excitations per site within the pair class (excluding input and output
sites) revealed that backbone and pair sites can be clearly divided into
{\it active} and {\it inactive} exciton carriers, respectively (see Supplementary Figure S5). 
In fact, backbone sites present
maximum excitation probabilities always larger than 0.25 while pair sites 
never more than 0.075 in 99\% of the cases.

Strikingly, removal of the pair from the original structures resulted in a
systematic efficiency loss of up to 0.32, which is rather surprising given that pairs
hardly serve as carrier of the excitation.  Efficiency loss is particularly 
severe for the most efficient realizations due to the sensitivity of perfect 
constructive interference against perturbations (Fig.~\ref{fig:pairs}b). 
Pair removal affects exciton transport in non-trivial
ways as shown by the quantum dynamics of the most efficient structure before
and after pair-removal (see Fig.~\ref{fig:pairs}c-d). In the modified
structure, we found unbalanced transport along backbone sites
contrary to what was originally observed as well as an inability of the input
site to promptly transmit the excitation to the closest backbone site. 

\begin{figure}[t!]
  \includegraphics[width=60mm]{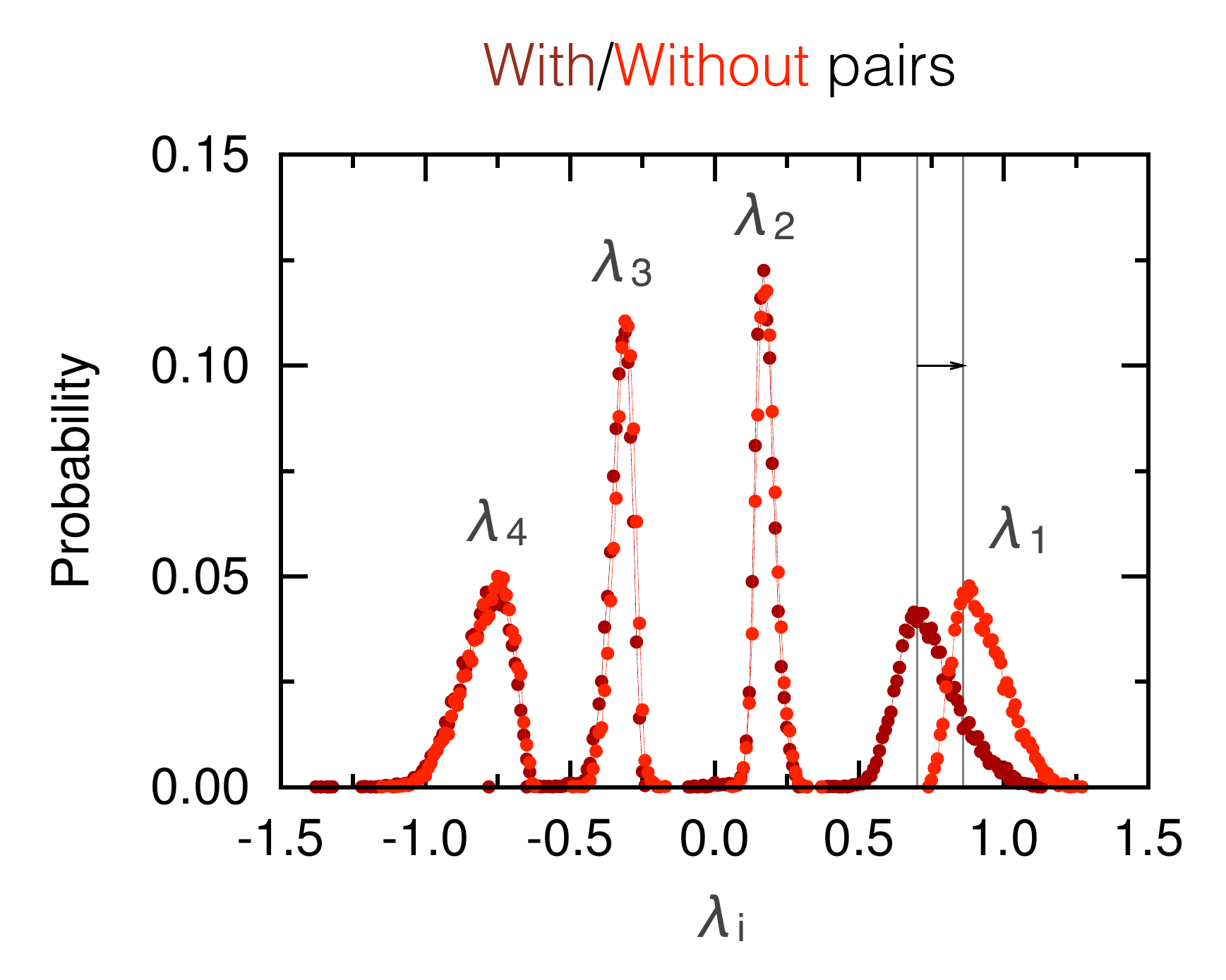} 
  \caption{Distribution of eigenvalues $\lambda$ in the presence or absence of interaction with the pair in the first cluster (in red and dark red respectively). 
In the presence of the pair the differences of the $\lambda_i$ are close to integer multiples of a fundamental frequency (see Supplementary Figure S6).}
  \label{fig:lambdashift}
\end{figure}

To investigate the mechanism behind the dynamical influence of the pair, we compared the distribution of the energy eigenvalues with and without the two pair sites. Given the weak interaction between the backbone and the pair, there are $N-2$ eigenstates with the excitation localized on the backbone; in the remaining two eigenstates the excitation is delocalized on the pair symmetrically (antisymmetrically), {\it i.e.} in the form of a triplet (singlet) state $|\pm\rangle=(|01\rangle\pm|10\rangle)/\sqrt{2}$. As depicted in Fig.~\ref{fig:lambdashift}, the interaction between the backbone and the pair sites results in a shift of the eigenfrequencies of the former $N-2$ eigenstates (denoted by $\mathit{\lambda}_i$ ($i=1,..,4$) in Fig.~\ref{fig:lambdashift} and Supplementary Figures S6 and S7, such that their differences  are close to integer multiples of a fundamental frequency. With this shift, the excitation is transferred to the output site essentially perfectly after one period of this fundamental frequency. The frequency shift results from a coupling between the former $N-2$ eigenstates and the triplet state; perturbatively it reads $|v|^2/\delta$, where $v$ is the interaction between backbone and the pair and $\delta$ is the interaction between the two pair sites, {\it i.e.} the eigenfrequency of the triplet. Since all coupling elements in \eqref{eq:ham} have the same distance-dependence ($\sim 1/r^3$), the energy shift is roughly independent under a change of the backbone-pair distance by a factor of $\alpha$ and a simultaneous change of the pair-size by a factor $\alpha^2$. This expectation is explicitly verified in the Supplementary Figure S4d, where a broad parabolic plateau of high efficiency is clearly discernible (yellow area). Only for small backbone-pair distances in the non-perturbative regime the plateau breaks down (Supplementary Figure S7). The robustness with respect to perturbations and the resulting statistical significance of the pair mechanism can also readily be deduced from the perturbative mechanism. Since  a loss of efficiency due to a change of distance between the backbone and the pair can be compensated by a change in the distance between the pair sites, there is not a unique or discrete set of optimal spatial configurations, which one would have otherwise expected for a mechanism based on constructive interference of many paths. Instead, what is found is a higher-dimensional manyfold of optimal configurations.

\begin{figure}[t]
  \includegraphics[width=60mm]{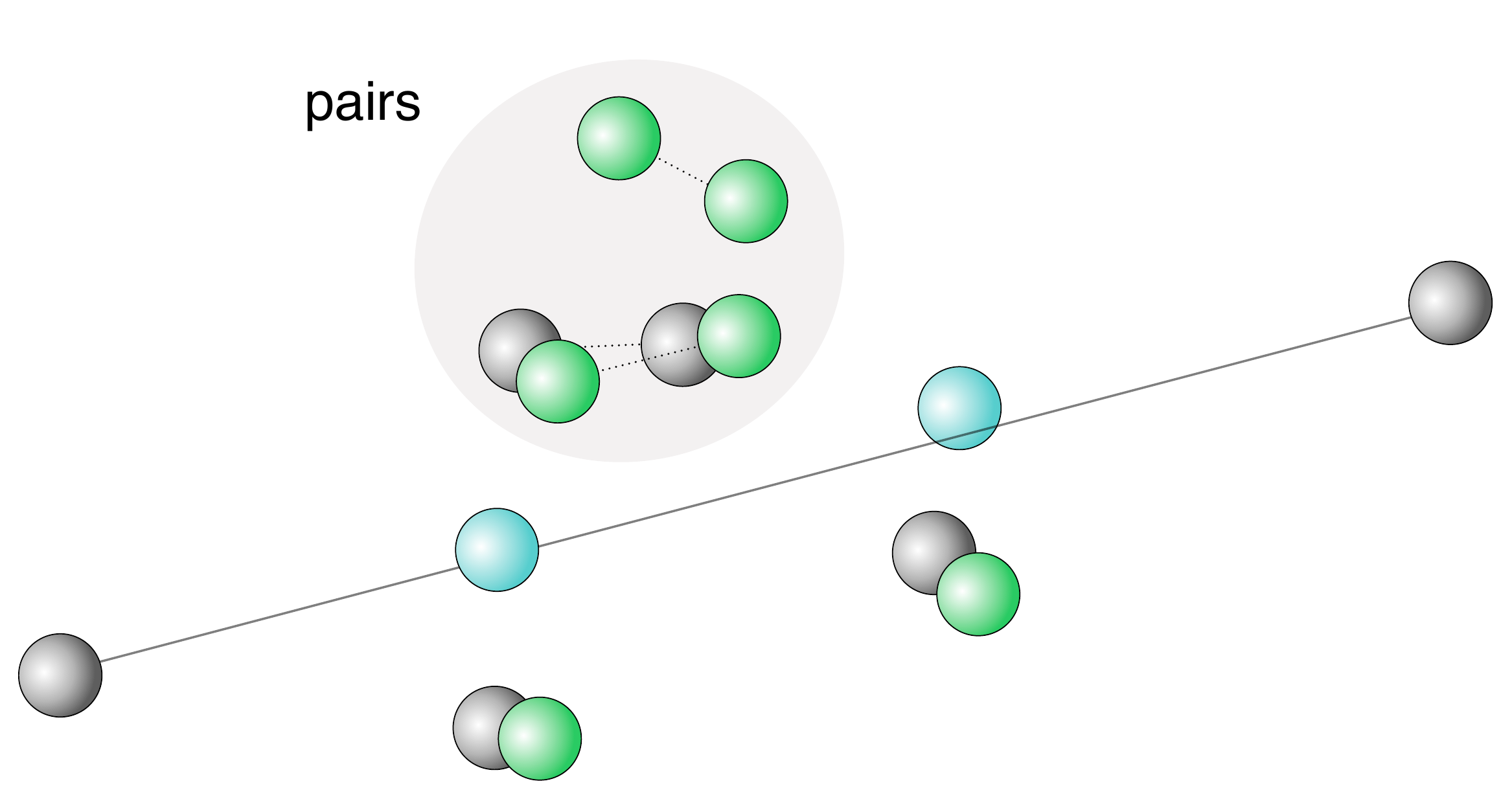} 
  \caption{The most efficient structures in the case $N=4$ ($\epsilon=0.922$), 
  $N=6$ ($\epsilon=0.998$) and $N=8$ ($\epsilon=0.993$).  The three cases are shown in light blue, gray and light green, respectively. The
  gray area highlights the presence of pairs for $N=6$ and $N=8$.}
  \label{fig:N4N6N8}
\end{figure}

From a geometrical point of view the pair class provides attributes commonly
observed in models with different number of sites.  For $N=4$
the ensemble of structures with $\epsilon > 0.9$ organized on a line with inter-site
spacings very similar to the ones found in the backbone of the pair class.
This is shown in Fig.~\ref{fig:N4N6N8} by superimposing the most efficient
structures for $N=4$ and $N=6$ (light blue and gray spheres, respectively).
However, the maximum efficiency for $N=4$ is of only 0.922 while the presence
of pair sites lead to a maximum efficiency of 0.998 which is also the best
value achieved within the whole sample of 100 millions structures. The general
role of pairs is confirmed by the $N=8$ case. In this model roughly half of the
whole sample with $\epsilon>0.9$ presented a modular structure with four out of eight
sites organized in 2 pairs with the remaining sites perfectly overimposed to the
backbone of $N=6$ (green spheres in Fig.~\ref{fig:N4N6N8}). Such geometry
clearly represents a generalization of the pair class. The pair structure was
also observed with $N=7$, where the backbone was formed by five equidistant sites.
Altogether, these results provided strong evidence that pair sites play an important structural
role to tune up quantum efficiency, suggesting the idea that they represent a
general building strategy towards efficient transport in multi-body quantum
systems.

The presence of active and inactive carriers of the transport efficiency in the
pair class pointed out a potentially intriguing scheme to rationalize the quantum
dynamics of all classes (Supplementary Figure S5).  Taking 0.075 as a threshold to define a site as
inactive, the pair class is characterized by a total of four active sites
including the input and output (see probability distribution of active sites in
the inset of Fig.~\ref{fig:dynamics}a). However, for the inline and sparse classes 
no clear separation into number of active sites was found (inset plots in panel b and c of Fig.~\ref{fig:dynamics}, respectively, see also Supplementary Figure S5).
Moreover, when this concept was used for the analysis of the efficiency loss
upon random displacements by grouping structures according to the number of
active sites, ambiguities raised between the five and six active sites groups
due to strong overlaps (see Supplementary Figure S8).
These results indicated that the active site concept alone is not sufficient to
define structural groups with similar dynamics, reiterating the idea that
advanced techniques for structural comparisons, like the one performed here, are
necessary to unravel the connection between structure and quantum dynamics.

\subsection{The incoherent case}
All the presented analysis was performed with perfectly coherent dynamics. 
The impact of noise on our analysis was investigated by considering environmental models leading to both Markovian and non-Markovian dynamics (see Supplementary Note 2 for
details). Within each case, addition of noise consistently decreased the efficiency with no specific distinction among the three classes,
 most importantly without affecting the response to random displacement classification identified with coherent
dynamics (see Supplementary Figures S9 and S10).
The purely destructive effect of noise is consistent with
the type of analysis performed, which focused on outstandingly fast and
efficient transport made possible only by constructive interference \cite{Zech2012}.
This represents a different scenario with respect to problems where environmental noise can have a beneficial effect \cite{Plenio2008,Mohseni2008,chin2013,Chin2010}.

\section{Discussion}

In conclusion, our work provided strong evidence for the emergence of non-trivial
characteristic structural motifs leading to high quantum transport
efficiencies. Specifically, the identification of site pairs that do not
participate directly to the exciton dynamics results in a general strategy to both
enhance quantum efficiency and to make structures robust against geometric
perturbations.  Consequently, a design principle is presented, exploiting
enhanced quantum transport in cases where perfect interferometric stability is
impossible. Gaining control on this problem is of paramount importance towards
the rational design of technologies making use of constructive interference. 
The analysis of the pair class led to the identification of a modular arrangement of the dynamics within these structures: an active 4-sites backbone accompanied with an inactive pair. Such a modular active/inactive arrangement of the excitation dynamics has been also identified in natural light harvesting complexes like FMO \cite{Brixner2005}. The seven chromophores can be divided in two weakly interacting sets, corresponding to an approximately block-like structure of the Hamiltonian \cite{Adolphs2006a}: chromophores 1 and 2 are strongly coupled to each other, and weakly with the remaining sites excluding the output number 3. This entails that, upon excitation of chromophore 6 (one of the two supposed entry points of the excitation from the antenna), the chromophores 1 and 2 seem to be effectively decoupled from the dynamics, just as the pairs in our system, but might still analogously influence the dynamics.

There are however a number of differences that must be kept in mind when comparing the results of the present work with natural light harvesting complexes. At a fundamental level, the ratio behind the definition of efficient transport follows from a different perspective, namely transfer being {\it lossless} in the FMO and {\it fast} in this work. Furthermore, care is needed when comparing geometries of a network of point-like entities with distance based coupling with a much more complicated pigment-protein complex.

In light of these differences a straightforward identification of geometric arrangement akin to the pair class in realistic natural systems should not be expected. In other words, in the two systems we find a similar modular arrangement of the {\it dynamics}, which, due to the differences in the model, do not necessarily arise from similar {\it geometrical} arrangements.

\section{Methods}

\subsubsection{Network Creation}

Network links are put according to a similarity parameter.  The 
similarity parameter $S$ is calculated as:
\begin{equation} S^2 = \sum^n_{i=1}  d_i ^2/n \end{equation}
where $d_i$ is the distance between corresponding sites
and $i$ runs over all six sites.  The
sites are indistinguishable, thus all different permutations of the site
labels need to be performed (in number of $(N-2)!$, excluding input and output). In
addition, symmetry along the in-out axis and an additional mirror symmetry has
to be taken into consideration.  The measure $S$ between configurations A and B
is calculated as follows: keeping fixed the set of labels for A, the
labels of B are changed. For each of these $(N-2)!$ sets, 180 rotations of 2
degrees each of the configuration B are done around its in-out axis.  For each
rotation, a mirror reflection relative to the $x-y=0$ plane is also done. The
final value of $S$ is then the minimum value of $\sum^n_{i=1}  d_i ^2/n$ among
the $(N-2)! \times 180 \times 2$ possible combinations of labels, rotation and
mirror states. 

Only $S^2$ values below $0.0125 \ r^2_0$ are considered as links in the network. That
means, the average $d_i$ between sites of two superimposed
configurations must be smaller than $\sqrt{0.0125 \ r^2_0} \sim 0.11 \ r_0$. 
This value lies just above the tail of the pairwise distance distribution (inset Supplementary Figure S11a). Consequently, only
the most similar structures are linked together.  Lower values of the cut-off would generate a disconnected 
network, while values too close to the maximum of the distribution would put links between 
structures that are not very similar. A double check with another cut-off value of $S^2< 0.0138 \ r^2_0$ was
performed, giving results very similar to the ones shown in the main text (see Supplementary Figure S11b-c).
Although the change from $S^2 < 0.0125 \ r^2_0$ to $S^2 < 0.0138 \ r^2_0$ seems negligible, it is worth noticing that it corresponds
to an increase of the total number of links from $1.03 \cdot 10^6$ to $1.42 \cdot 10^6$, i.e. a considerable $40\%$.

\subsubsection{Clusterization Procedure}

In order to obtain different homogeneous classes of configurations, the Markov
cluster algorithm (MCL) was used \cite{gfeller2007,Van2008}. This algorithm is based on
the behavior of random walks on the network and consists of four
steps: (i) start with the transition matrix $C$ of the network, where
each column is normalized to 1; (ii) compute $C^{2}$; (iii) take
the {\it p}th power ($p>1$) of every element of $C^{2}$, normalizing afterward
each column and (iv) go back to step (ii).  After some iterations MCL
converges to $C_{\text{MCL}}$, where only few entries for each column are non-zero
(exactly only one non-zero entry per column). These entries give the clusters. 
The parameter {\it p} is related to the
granularity of the clustering process. High values of {\it p} generate several small homogeneous clusters. On the other hand, in the
limit of $p=1$, only one cluster is detected. In this work
$p=1.4$ was used, giving a very good signal-to-noise ratio (see Supplementary Figure S1).
Smaller values of $p$ gave similar results. 
For instance, $p=1.2$ splits the network into two clusters: the one corresponding to the pair class
described in the main text and the remaining two classes all together. This suggests that differences
between the clusters in inline and sparse classes appear at a finer
degree of granularity, while separation between these and the pair class
is more evident, requiring a lower parameter $p$ to be resolved.

\subsubsection{Structural superposition}

Figures \ref{fig:dynamics} and Supplementary Figures S2 and S3
are obtained as follows: for each cluster, the most connected
structure is taken as reference and all the others are superimposed to
that. For each of them the combination of labeling, rotation and mirror state
which minimizes the similarity parameter $S$ was taken (see {\it Network Creation} 
section). In order to reduce noise, the coordinates of the sites were averaged
with the ones from two other structures of the cluster taken at random. 
Structural rendering was done with VMD \cite{humphrey1996vmd}.

\section*{Contributions}
SM, FL, DPG, FM and FR designed the experiment and analyzed the data. SM, FL and DPG wrote the analysis codes. SM, FL, FM and FR wrote the paper.

\section*{Acknowledgments}

This work is supported by the Excellence Initiative of the German Federal and State Governments.
FM gratefully acknowledges financial support by the European Research Council within the project odycquent (259264).

\newpage
\clearpage

{\LARGE \bf Structure-dynamics relationship in coherent transport through disordered systems - Supplementary Information} 

\begin{figure*}[!h]
  \includegraphics[width=60mm]{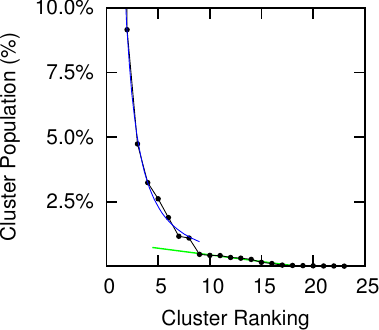}
  \caption*{ {\bf Supplementary Figure S1 |} Cluster populations. The populations of the clusters from 2 to 8
  are fitted by a power law function (blue line).  The first eight clusters represent cumulatively the 97\% of the whole sample. The rest of the clusters, here fitted by a green line, are considered as noise. }
  \label{fig:clustpop}
\end{figure*}

\begin{figure*}[!h]
  \includegraphics[width=180mm]{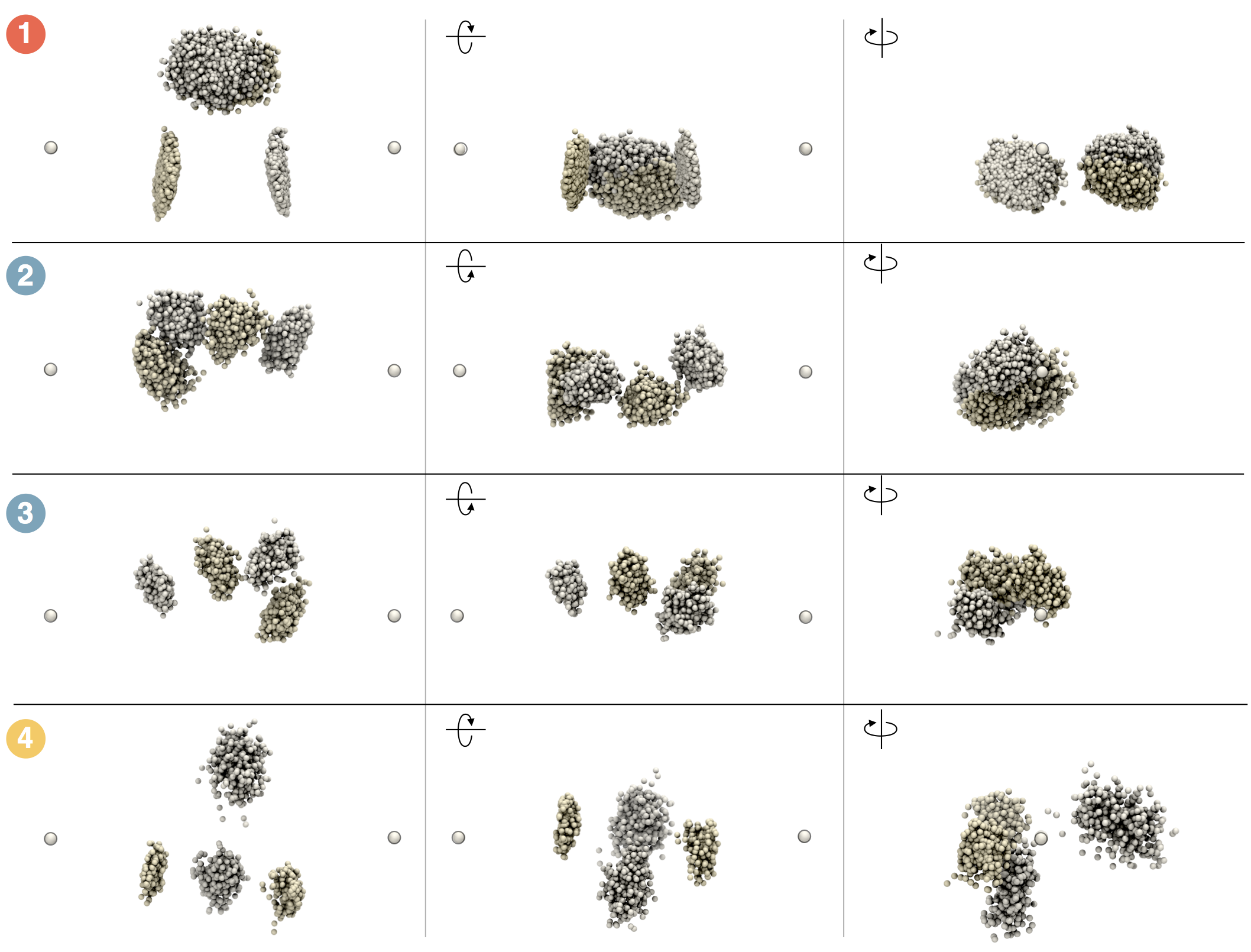}
  \caption*{{\bf Supplementary Figure S2 |}Superimposition of all structures belonging to clusters 1 to 4. The cluster number is colored according to the class of affiliation (red, blue and yellow for pair, inline and sparse classes, respectively). For clarity, structures are shown in three different orientations.}
  \label{fig:homogeneityclusters1to4}
\end{figure*}

\begin{figure*}[!h]
  \includegraphics[width=180mm]{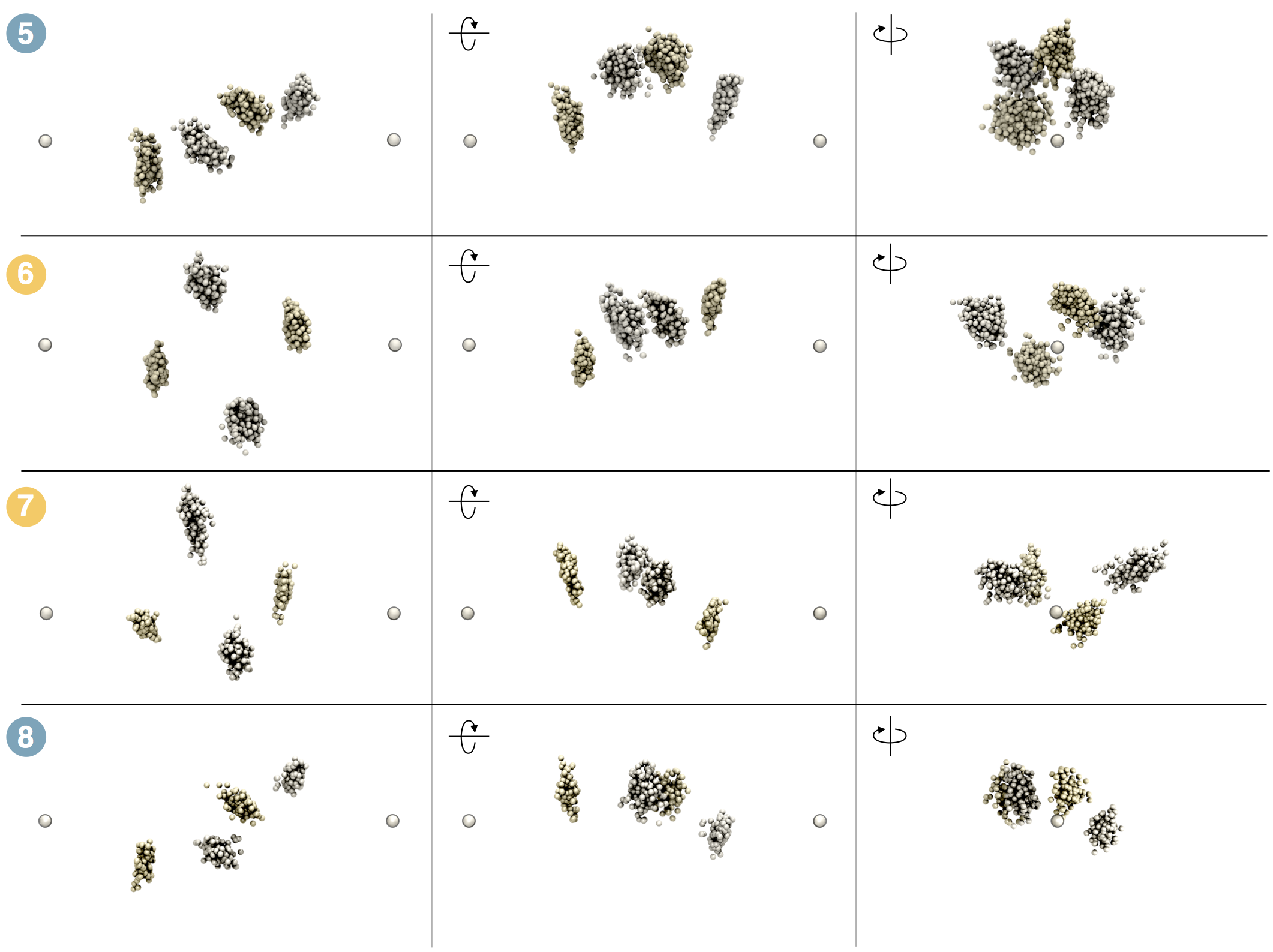}
  \caption*{{\bf Supplementary Figure S3 |}Superimposition of all structures belonging to clusters 5 to 8.
  }
  \label{fig:homogeneityclusters5to8}
\end{figure*}

\begin{figure*}[!h]
  \includegraphics[width=0.9\textwidth]{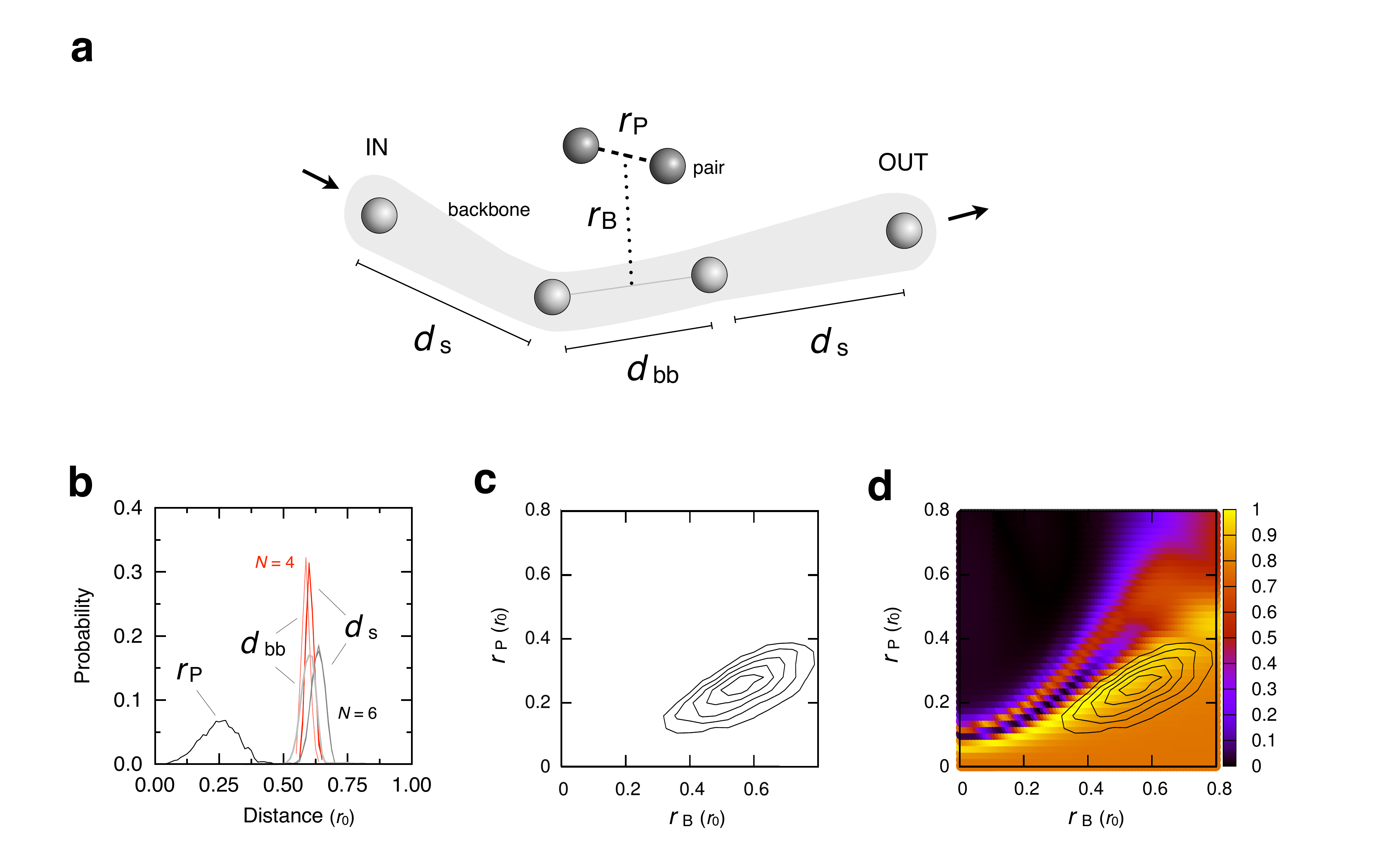}
\caption*{{\bf Supplementary Figure S4 |} Pairs definition in space. (a) Pairs of the first cluster can be localized by means of the intra pair distance $r_{\text{P}}$ and the distance between the pair and the backbone centers of mass $r_{\text{B}}$. Intra-backbone distances $d_{\text{s}}$ and $d_{\text{bb}}$ are also shown. (b) The distances $d_{\text{bb}}$ and $d_{\text{s}}$ show similar distributions in both cases ($0.60-0.64 \cdot r_0$ for $N=6$ and $0.59-0.61\cdot r_0$ for $N=4$), revealing a similar backbone geometry, while the intra pair distance $r_{\text{P}}$ is much smaller ($0.25 \cdot r_0$). (c) Density plot of the pair position from the entire class as a function of $r_{\text{P}}$ and $r_{\text{B}}$. A large portion of the configurational space is spanned by the pairs. (d) Efficiency as a function of $r_{\text{P}}$ and $r_{\text{B}}$ (backbone sites are the ones of the most efficient structure). Data was generated by moving the pair sites along $r_{\text{P}}$ and $r_{\text{B}}$ while keeping the backbone sites fixed. A 1D manifold emerges where the pair can move without affecting the efficiency (yellow region with $\epsilon>0.9$). The small differences between backbones belonging to different structures (panel b) and the configuration of the pairs in space across all structures (panel c) suggest how the plot in panel d approximately holds regardless of the reference backbone. Therefore, the majority of the pairs can then be located within this 1D manifold, providing a first argument in favour of the increased robustness of structures within the pair class}
  \label{fig:comparisonNtot4}
\end{figure*}

\begin{figure*}[!h]
  \includegraphics[width=70mm]{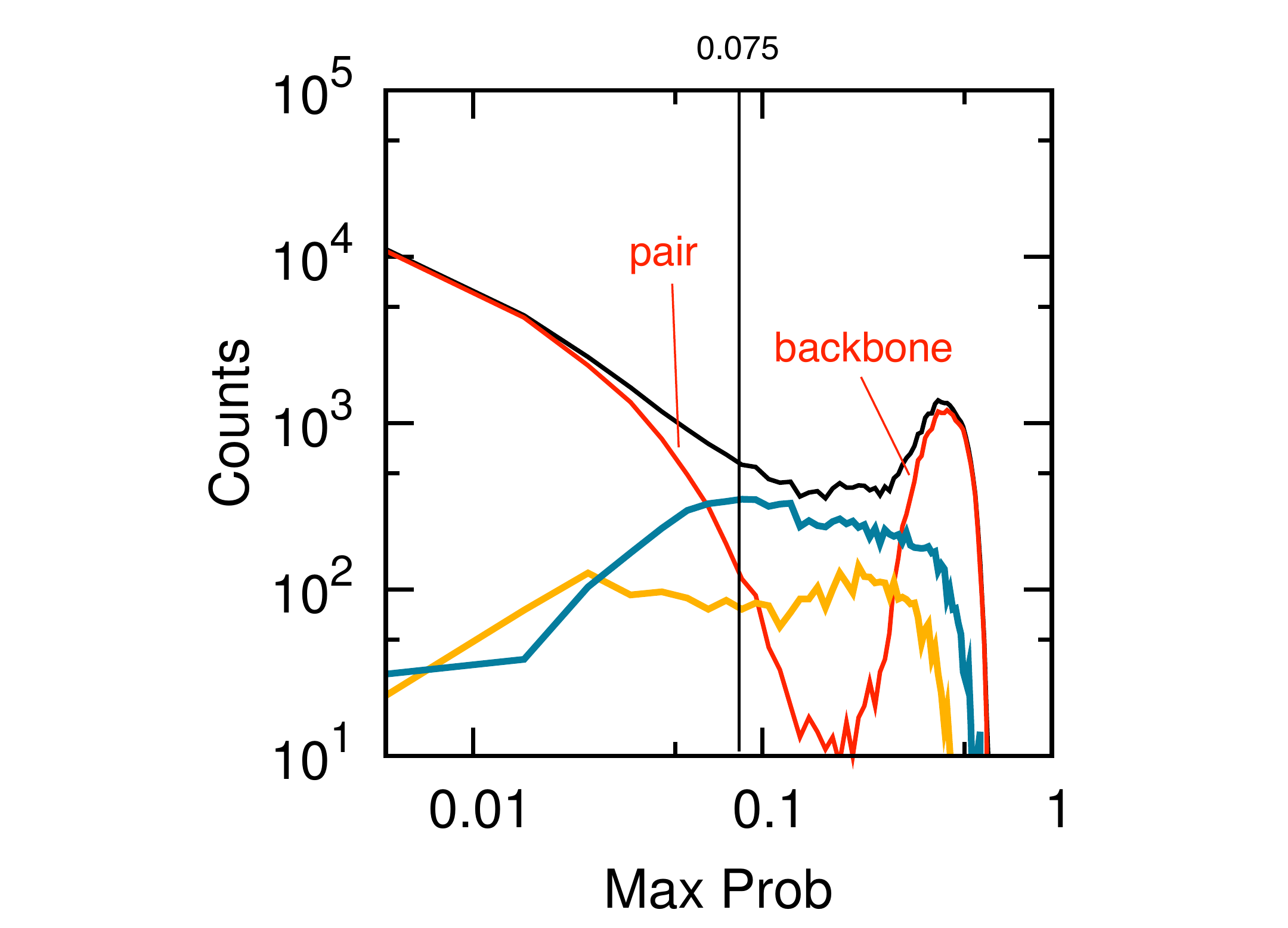}
  \caption*{{\bf Supplementary Figure S5 |}Distribution of the maximum of the site excitation probability within time $\tau$ (black line, in and out sites are not included).  The distribution of the pair class reveals two trends (red line): roughly half of the sites never gets excited more than 0.15 (the 99\% of which never more than 0.075), the other half has the maximum of excitation above 0.25. On the other hand, sites belonging to the sparse (yellow) and inline classes (blue) present no evident separations.
  }
  \label{fig:activeinactive}
\end{figure*}

\begin{figure*}[!h]
\includegraphics[width=0.75\textwidth]{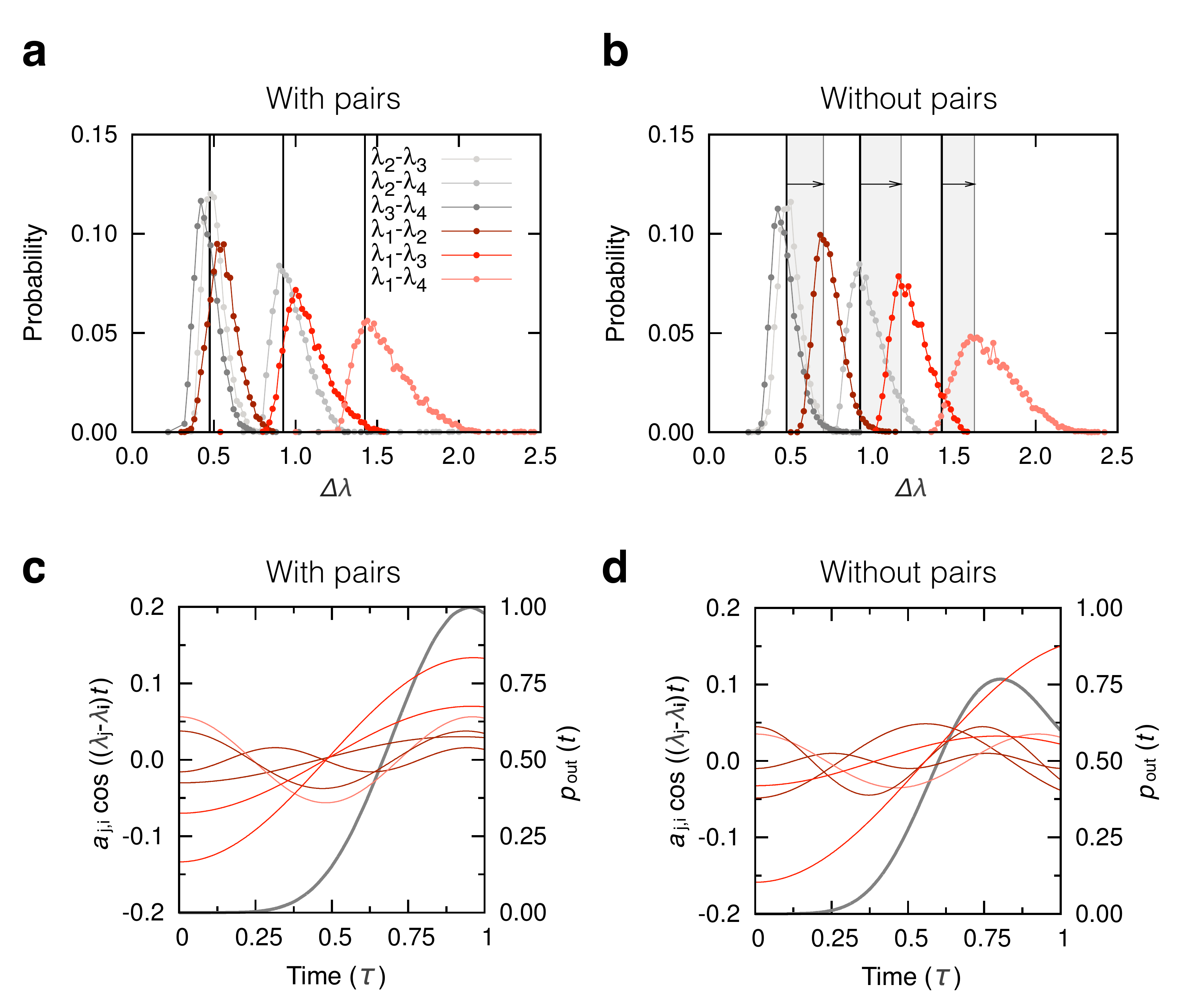} 
\caption*{{\bf Supplementary Figure S6 |}Pair removal causes $\lambda_1$ shifting and anharmonicity. (Top) Frequencies relative to the backbone group with (a) and without (b) the pair. The shift of the first eigenvalue causes anharmonicity of three frequencies (b), which are not anymore multiple of a constant as in case a. (Bottom) Example of the behavior of the 6 time oscillating contributions to $|\langle \text{in} |\mathrm{e}^{iHt}| \text{out} \rangle|^2$ with (c) and without (d) the pair. The anharmonicity induced by the pair removal results in the contributions having the maximums at different times within the $[0,\tau]$ time window, thus in an efficiency loss. The probability in the output site is shown in grey.}
\label{fig:deltae}
\end{figure*}

\begin{figure*}[!h]
\includegraphics[width=0.75\textwidth]{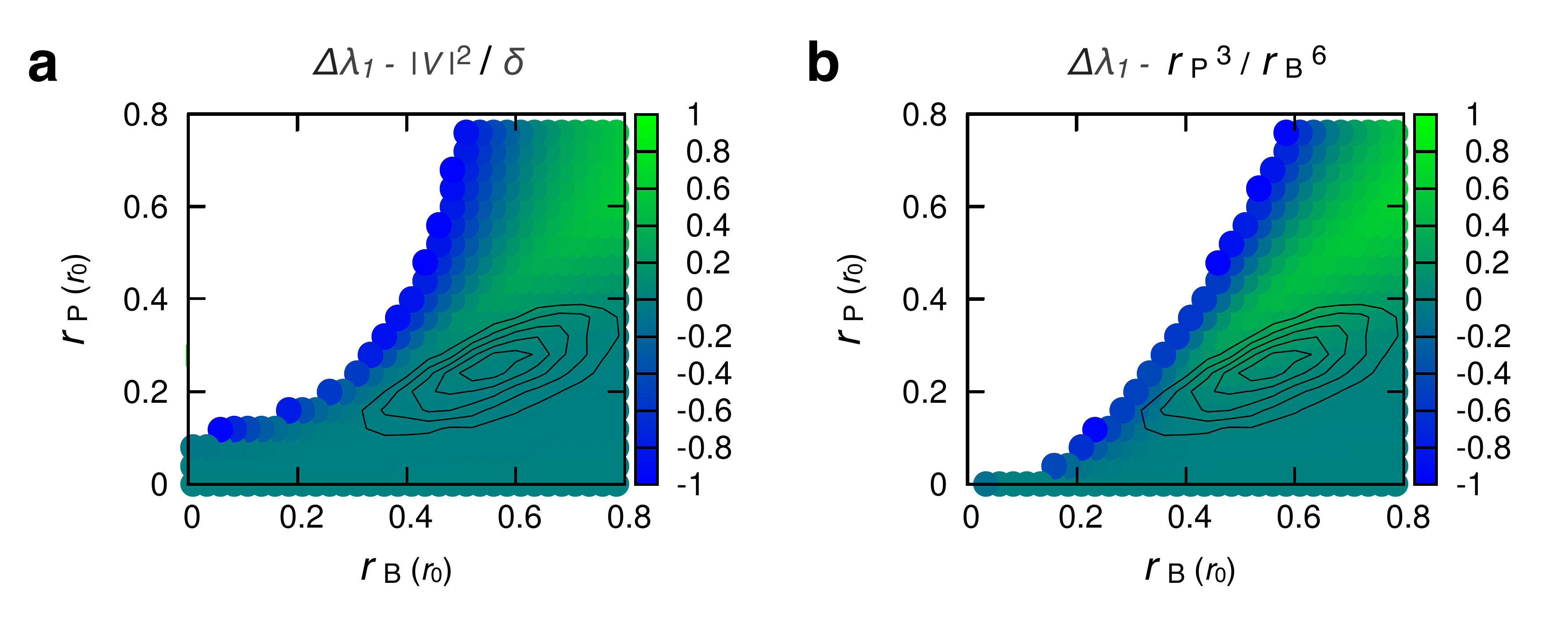}
\caption*{{\bf Supplementary Figure S7 |}Variation of the first eigenvalue $\lambda_1$ in presence or absence of the pair as compared to $\frac{|v|^2}{\delta}$ (a) and $\frac{r_{\text{P}}^3}{r_{\text{B}}^6}$ (b) (values are expressed as fraction of $\lambda_1$, and differences in absolute value higher than 1 are shown in white). The two approximations hold well in the region of $\{r_{\text{P}},r_{\text{B}}\}$ space where our structures lie (black contour lines; coordinates are defined in the caption of Supplementary Figure S4). }
\label{fig:deltalambda1}
\end{figure*}

\begin{figure*}[!h]
\includegraphics[width=60mm]{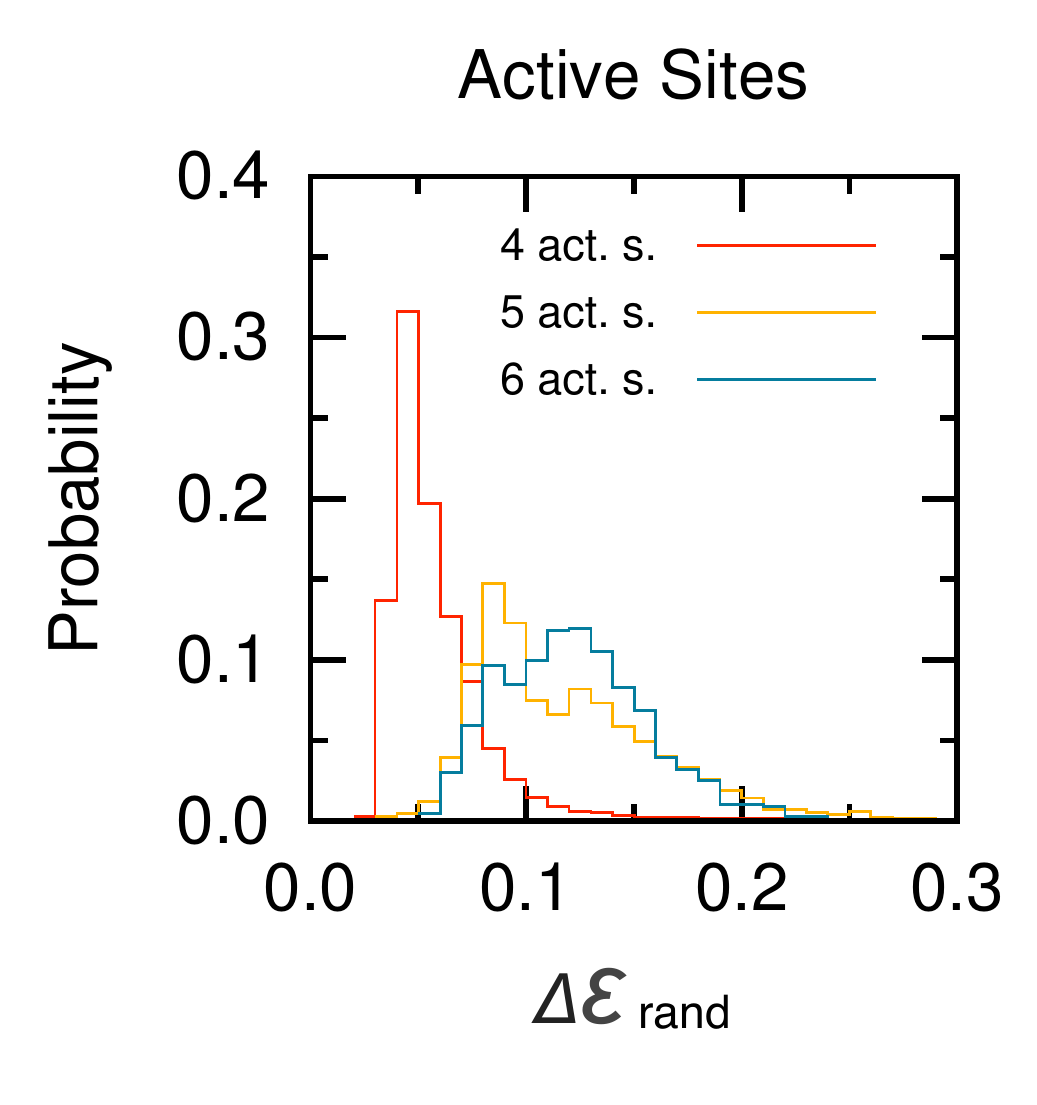}
\caption*{{\bf Supplementary Figure S8 |}Efficiency loss upon random displacements (see main text) according to the number of active sites. Structures with 4,5 and 6 active sites are shown in red, yellow and blue respectively. This classification is not able to resolve the three different responses to noise presented in Fig. 1b of the main text.}
\label{fig:randomactivesite}
\end{figure*}

\begin{figure*}[!h]
  \includegraphics[width=140mm]{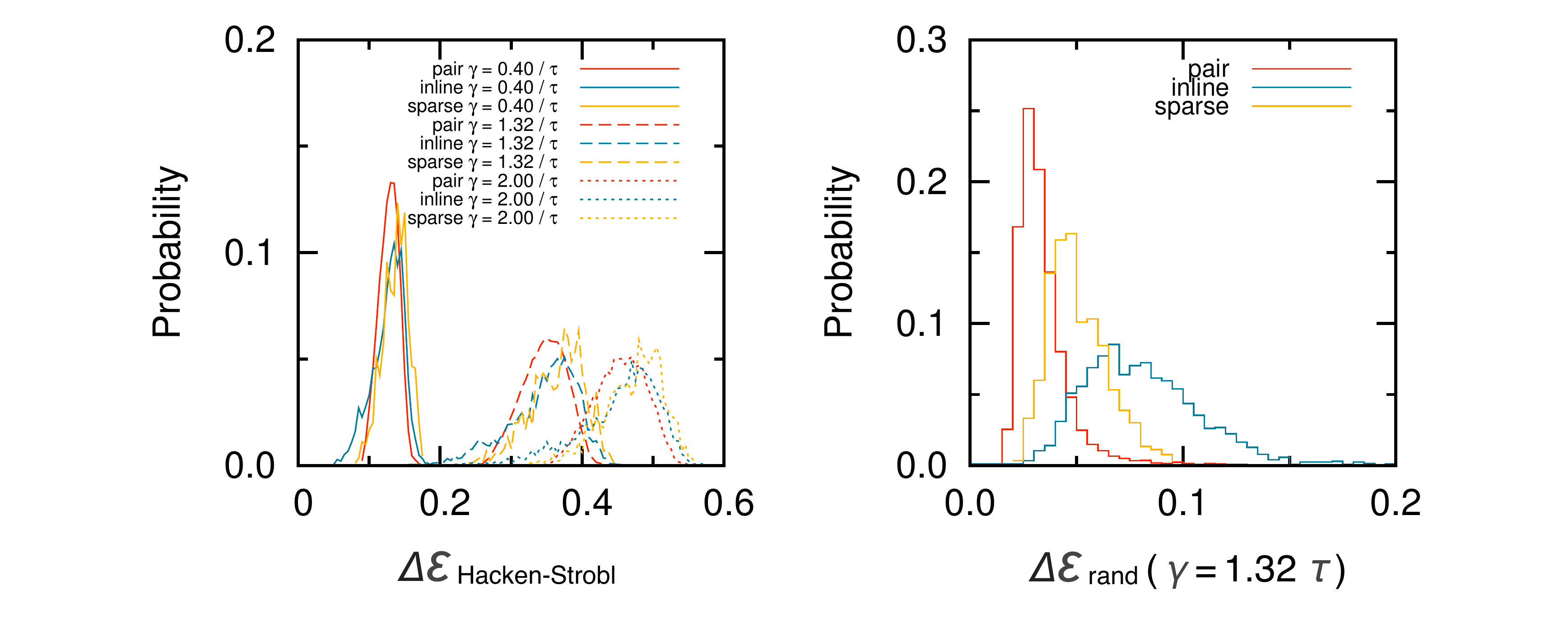}
  \caption*{{\bf Supplementary Figure S9 |}Efficiency loss upon the introduction of noise in the Haken-Strobl model. (left) The three classes are shown in red, yellow and blue for
  the pair, sparse and inline classes, respectively. Three different values of $\gamma$ 
  were used ($0.40$, $1.32$ and $2.00 / \tau$), here presented in solid, dashed and dotted lines,
  respectively. No evident difference in efficiency loss subsisted between the classes.
  (right) Efficiency loss upon random displacement in the presence of Haken-Strobl noise ($\gamma=1.32/\tau$).
  }
  \label{fig:dephasing}
\end{figure*}

\begin{figure*}[!h]
\includegraphics[width=0.75\textwidth]{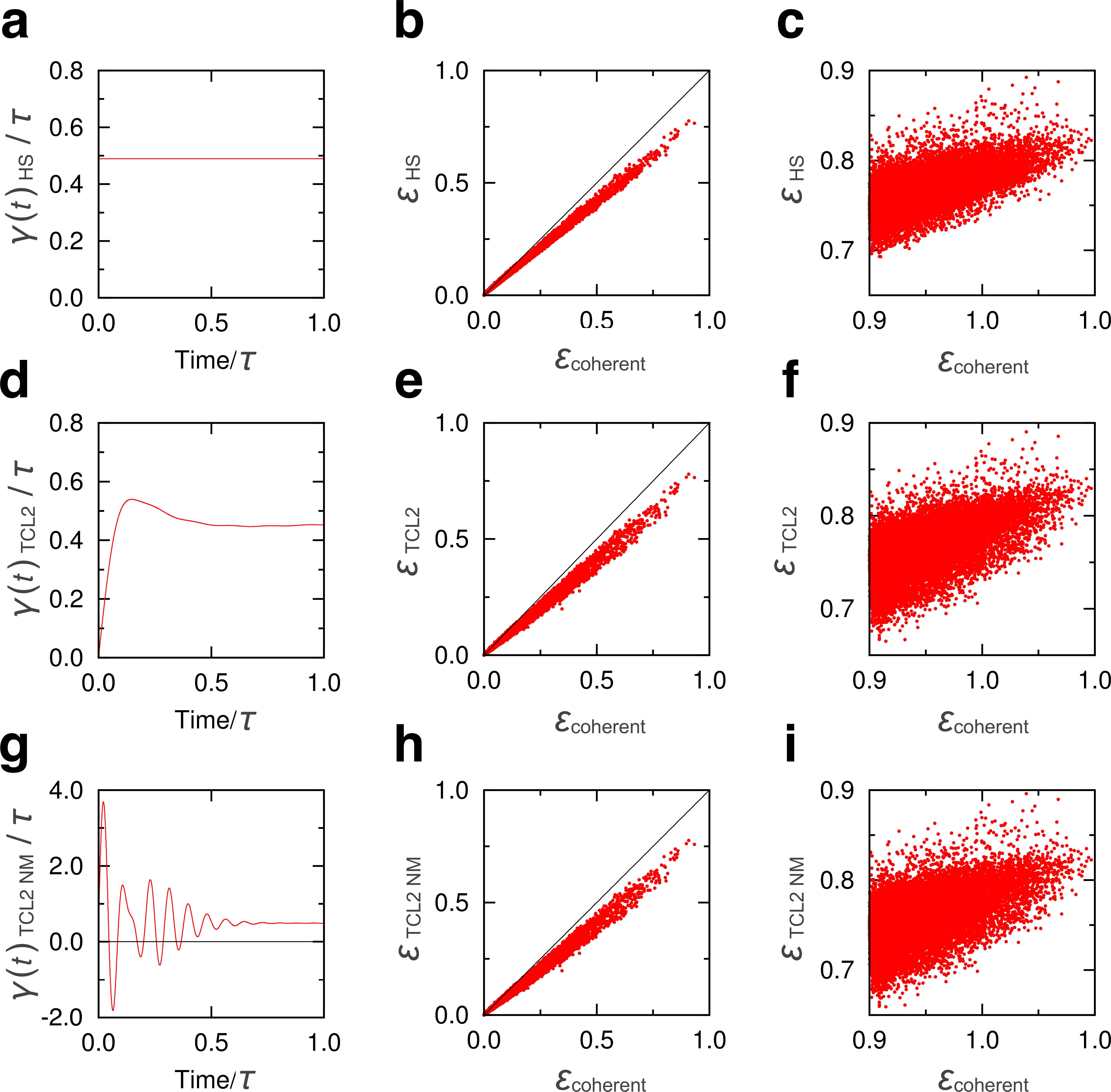}
\caption*{{\bf Supplementary Figure S10 |}Efficiency in non-coherent cases. Correlation between the perfectly coherent model and Haken-Strobl (a-c), time-dependent TCL2 (d-f) and non-Markovian TCL2 (g-i) models for the most efficient structures (right column) and 25.000 structures taken at random (center column). The different $\gamma(t)$ are shown on the three left panels.}
\label{fig:noise}
\end{figure*}

\begin{figure*}[!h]
\includegraphics[width=150mm]{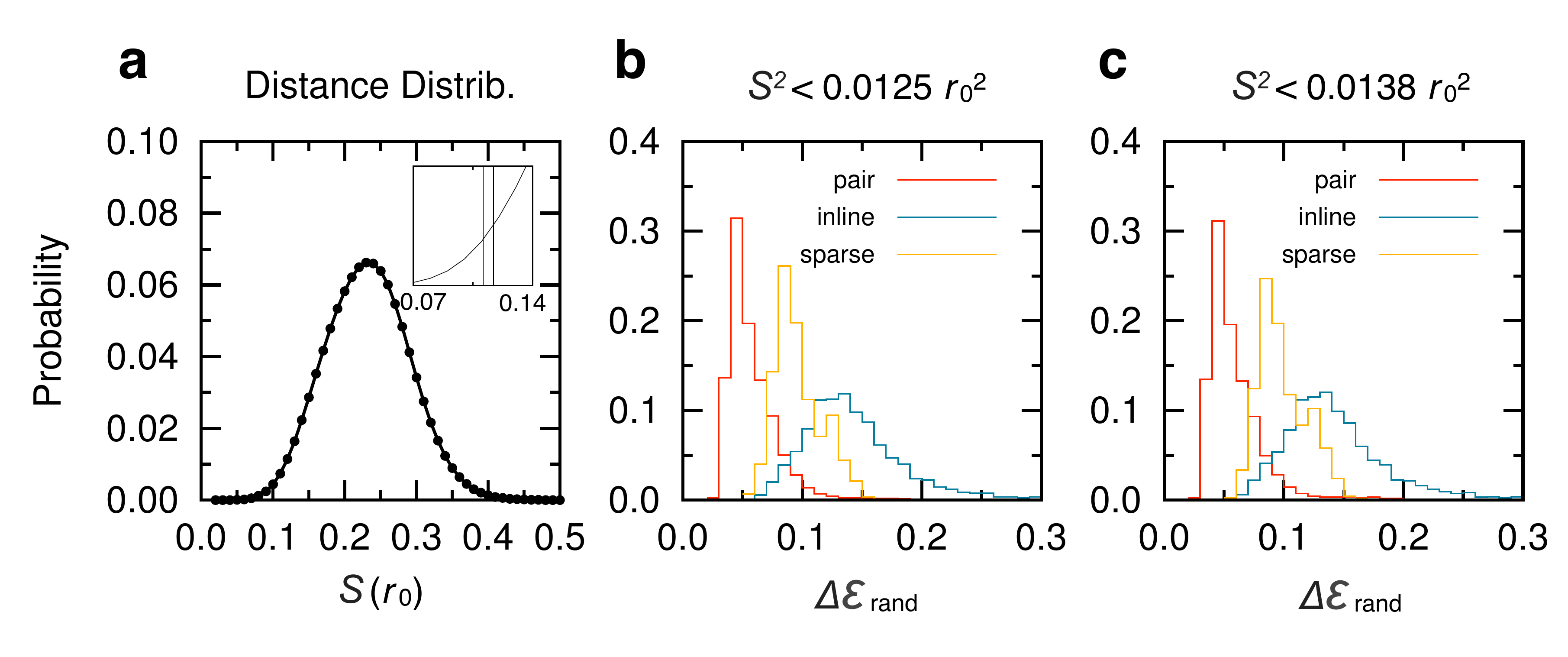}
\caption{{\bf Supplementary Figure S11 |}Distance cut-off. (a) Pairwise distance distribution between any-two structures for $N=6$ and $\epsilon>0.9$. The chosen cut-offs are shown as vertical lines in the inset. (b-c) Efficiency loss upon random displacements (see main text) according to the class partitioning proposed in the work with the two cut-offs. Increasing the number of links of the network by a $\sim 40\%$ (from b to c) does not affect the general behavior, suggesting that our results are robust for different values of the similarity cut-off.}
\label{fig:distancecutof}
\end{figure*}

\begin{figure*}[!h]
\includegraphics[width=0.3\textwidth]{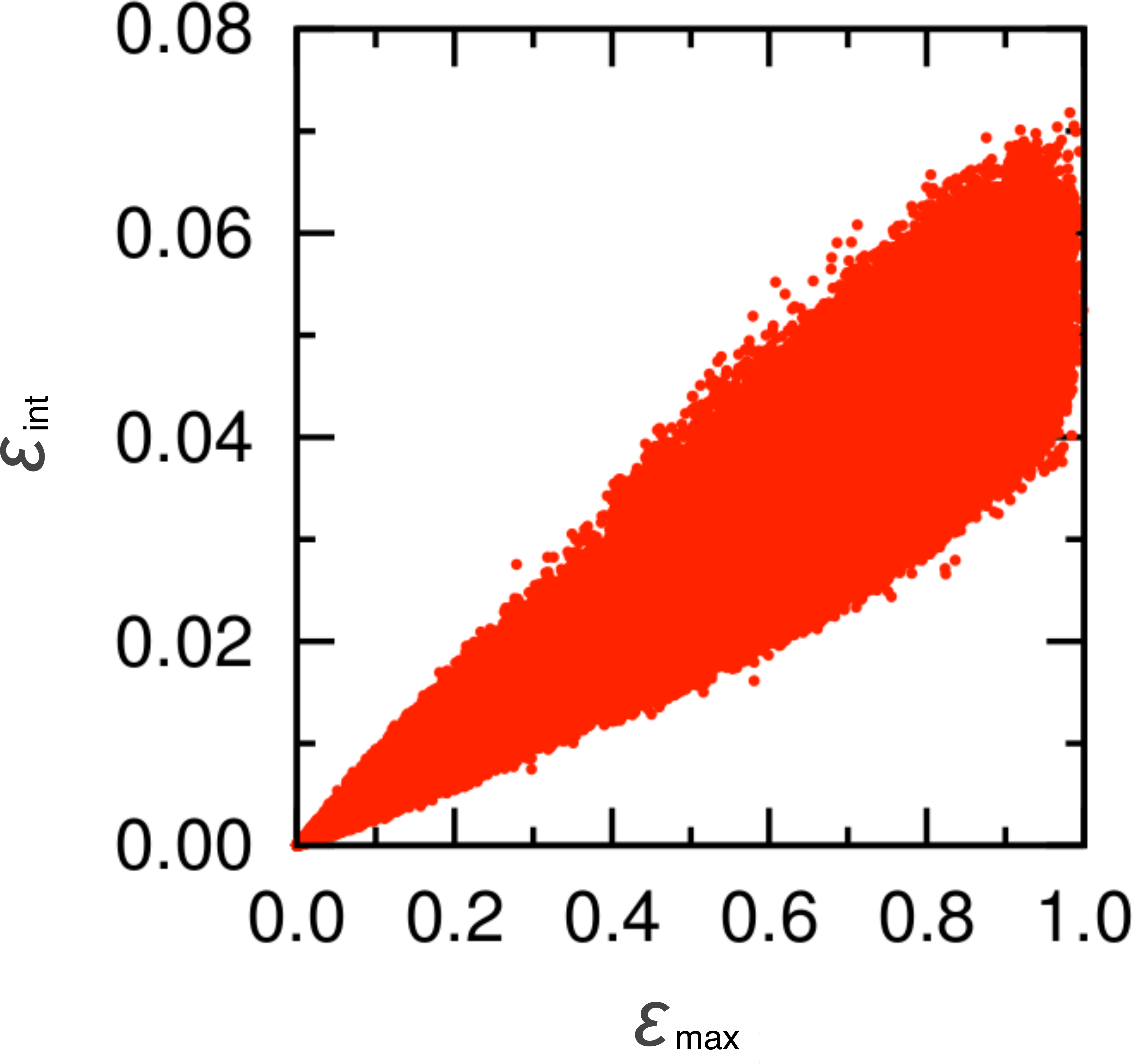}
\caption*{{\bf Supplementary Figure S12 |}Correlation between two different efficiency definition (see Supplementary Note 1). It is important to note that the $\epsilon_{\text{int}}$ definition is usually used with a sink and a sink rate, which is used to renormalize the values. In our case a sink is not present. This is the reason why the values of efficiency $\epsilon_{\text{int}}$ are {\it not} normalized.}
\label{fig:integrationmaxeff}
\end{figure*}

\begin{figure*}[!h]
\includegraphics[width=0.45\textwidth]{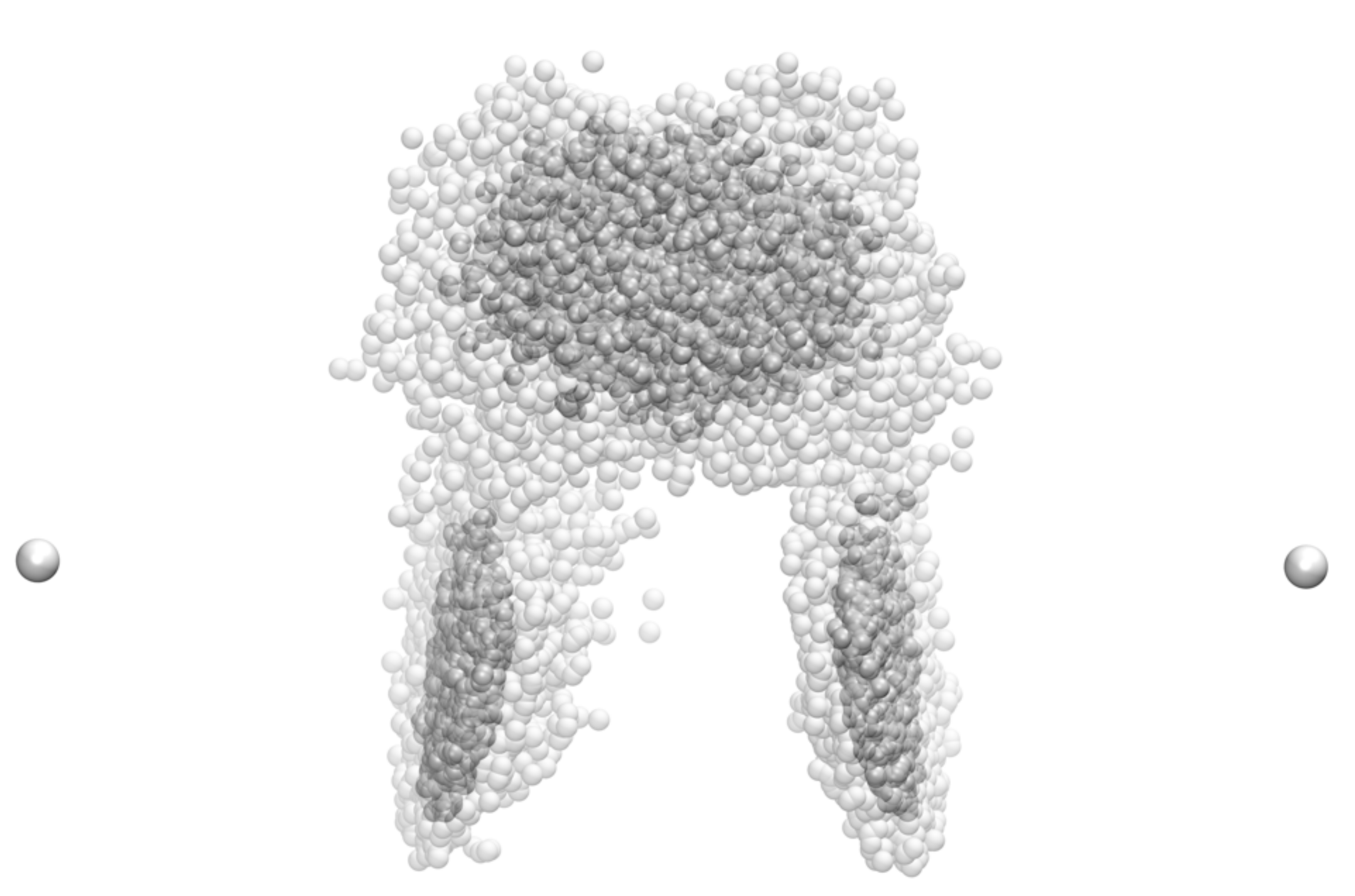} 
\caption*{{\bf Supplementary Figure S13 |} Structural characterization of the pair class for different values of the time window in the efficiency definition. Configurations obtained with a time window of $2 \tau$ (transparent) are superimposed on the ones relative to $1 \tau$ (darker). Even doubling the propagation time the same pair localization in the space was found.}
\label{fig:duetau}
\end{figure*}

\clearpage
\newpage

\subsubsection{{\bf Supplementary Note 1 | Efficiency definition and time interval}}

The most commonly employed figure of merit to assess transport properties is based on an integral of the probability to find the excitation at the output site over very long times (often infinity) [21,23]. Such a definition is connected to the fact that, especially in natural light harvesting complexes, energy transfer is considered {\it efficient} if it is {\it lossless}. This is a consequence of the different orders of magnitude that separate the exciton's lifetime and the transport time in these systems. With the efficiency defined in this way, one loses the possibility to distinguish between a slow or fast transport which, in turn, makes it harder to directly relate transport properties to the presence of quantum interference. For this reason we require that {\it efficient} transport must also be {\it fast}, by demanding that the excitation is delivered to the output site within a time-window $\tau$. Our efficiency is then defined as $\epsilon_{\text{max}}=\mathrm{max}_{t \in [0,\tau]} \{p_{\text{out}}(t)\}$ [13] which, by a suitably small enough choice of $\tau$, ensures that we study only ultra-fast transport that necessarily arise from constructive interference. For the sake of completeness we have performed a comparison between the values obtained according to our definition and an alternative $\epsilon_{\text{int}}=1/\tau \int_0^\tau p_{\text{out}}(t)dt$; a correlation between the outcomes is shown in Supplementary Figure S12, in agreement with [13].

Discussing the choice of the time interval $\tau$, it is important to note that the system we consider is not dissipative. This implies that the excitation will periodically oscillate back and forth between input and output for long times. If the time window is doubled the excitation returns to the input site and in some structures $p_{\text{out}}$ will even have more than one maximum. The parameter $\tau$ was therefore chosen in order to make sure the analysis is restricted to the first oscillation from input to output, so that there can be only one maximum of $p_{\text{out}}$ [13]. This again is in the spirit of targeting only fast transport. Nevertheless, to prove the robustness of our conclusions, we have reconstructed the network with time windows of $2 \ \tau$ ($\epsilon_{\text{max}}$ as efficiency definition). We found roughly twice as many efficient structures, but qualitatively the same geometries. Interestingly, the first cluster (with the 73\% of total population, the same as before) features the same geometrical motif with a very similar pair localization (Supplementary Figure S13). These results reinforce the idea that pair sites provide harmonic behavior, a result which is independent from the length of the time interval $\tau$.

\subsubsection{{\bf Supplementary Note 2 |Dynamics with noise}}

Loss of coherence in the system was introduced via the addition of an incoherent term to the dynamics of the quantum state. We investigated environmental effects modeled through a master equation of the form
\begin{equation}
\dot{\rho}(t)=-i[\rho(t),H]+\gamma(t) \sum_{k} \left( A_k \rho(t) A_k^{\dagger} - \frac{1}{2} \{ A_k^{\dagger} A_k ,  \rho(t) \} \right),
\label{eq:TCL2}
\end{equation}
where $A_k = | k \rangle \langle k |$ and the state $|k \rangle$ corresponds to having k-th site excited and all the others are in the ground state. This is obtained with the TCL method at the second order (TCL2) [26].

The first scenario we considered is the Haken-Strobl model [27,28] which is obtained from equation (\ref{eq:TCL2}) with a constant rate $\gamma(t)=\gamma$. The presence of noise leads to a systematic loss of efficiency, which is depicted in Supplementary Figure S9a for three values of the rate $\gamma=0.4/\tau,\,1.32/\tau,\,2/\tau$. No difference in efficiency loss has been detected between the three classes for any value of the rate. The response to geometrical perturbation of the structures remains furthermore invariant in presence of noise: the three classes emerge analogously as in the perfectly coherent case (Supplementary Figure S9b).

In order to investigate whether different noise models might lead to different conclusions, we have considered two further scenarios in the TCL2 framework, and compared them with what we obtained previously. We have evolved all the most efficient structures and a small sample of 25.000 random structures according to eq.~(\ref{eq:TCL2}) with different choices of the function $\gamma(t)$. The shape of the function $\gamma(t)$, the comparison of the perfectly coherent efficiency against the noisy one for the random sample and the most efficient structures are shown in the first, second and third columns of Supplementary Figure S10, respectively.

In Supplementary Figure S10a-c the resulting efficiencies under the influence of the Haken-Strobl model with a constant rate given by $\gamma=0.5/\tau$ are shown. We then considered a time-varying positive rate $\gamma(t)$ given by [29]
\begin{equation}
\gamma(t)=2 \int_0^{\infty} d \tilde{\omega} J(\tilde{\omega}) \coth \left( \frac{\hbar \tilde{\omega}}{2 k_\text{B} T} \right) \frac{\sin(\tilde{\omega} t)}{\tilde{\omega}}
\end{equation}
where
\begin{equation}
J(\omega)=\frac{\lambda}{\hbar \omega_\text{c}} \omega \exp \left( - \frac{\omega}{\omega_\text{c}} \right)
\end{equation}
is the Ohmic spectral density, $\omega_\text{c} =30$ cm$^{-1}$, $T=10$ K and $\lambda$ is set in order to converge to the constant Haken-Strobl rate $\gamma(+ \infty) = 0.5 / \tau$ to recover the previous models. Results of this time-dependent TCL2 approach are shown in Supplementary Figure S10d-f. Until now we have considered Markovian evolution for our noisy system. Non-Markovian effects are now introduced by considering a rate which gets negative for some times [29]:
\begin{equation}
\gamma(\omega,t)=2 \int_0^{\infty} d \tilde{\omega} J(\tilde{\omega}) \times
\end{equation}
\begin{equation*}
\times \left[ n(\tilde{\omega}) \frac{\sin((\omega+\tilde{\omega})t)}{\omega +\tilde{\omega}} + (n(\tilde{\omega})+1) \frac{\sin((\omega-\tilde{\omega})t)}{\omega - \tilde{\omega}} \right],
\end{equation*}
where $J(\omega)$ is again the Ohmic spectral density and
\begin{equation}
n(\omega)= \frac{\hbar \omega^3}{4 \pi^3 c^3} \frac{1}{\exp ({\frac{\hbar \omega}{k_\text{B} T}})-1} \; .
\end{equation}
\noindent A single channel $\omega=150$ cm$^{-1}$ was considered, with $\lambda=30$ cm$^{-1}$, $\omega_\text{c} =10$ cm$^{-1}$, $T=10$ K (Supplementary Figure S10g-i, with the other constants set to 1).

As noticeable from Supplementary Figure S10 both the random sample and the most efficient structures show how the efficiency loss is proportional to the original perfectly coherent efficiency in all these models. The resulting efficiencies are on the same line with the ones obtained within the Haken-Strobl framework (Supplementary Figure S9), showing that our results are independent from the specific noise model considered.\\[.9mm]

\newpage
\newpage

\section{Supplementary References}

[26] Breuer, H.~P. and Francesco, P. \emph{{The Theory of Open Quantum Systems}}. Oxford University Press Inc., New York, (2002).

[27] Haken, H. and Strobl, G. {An exactly solvable model for coherent and incoherent exciton motion}. \emph{Zeitschrift fuer Physik}, 262\penalty0 (2):\penalty0 135--148, (1973).

[28] Rebentrost, P., Mohseni, M., Kassal, I., Lloyd, S. and Aspuru-Guzik, A. \newblock {Environment-assisted quantum transport}. \newblock \emph{New Journal of Physics}, 11\penalty0 (3):\penalty0 033003, (2009).

[29] Rebentrost, P., Chakraborty, R. and Aspuru-Guzik, A. \newblock {Non-Markovian quantum jumps in excitonic energy transfer.} \newblock \emph{The Journal of chemical physics}, 131\penalty0 (18):\penalty0 184102, (2009).

\end{document}